\documentstyle[graphicx,twocolumn,pre,aps]{revtex}
\newcommand{\outputfig}[3]
{\includegraphics[height=#2\linewidth,width=#3\linewidth]{#1}}

\begin{document}
\draft
\title{Tunneling Mechanism due to Chaos in a Complex Phase Space}
\author{T.Onishi, $\! \! ^{1}$ A.Shudo, $\! \! ^{1}$ K.S.Ikeda, 
$\! \! ^{2}$ and K.Takahashi $\! \! ^{3}$}
\address{
$^{1}$Department of Physics, Tokyo Metropolitan 
University, Minami-Ohsawa, Hachioji 192-0397, Japan \\
$^{2}$Faculty of Science and Engineering, 
Ritsumeikan University, Noji-cho 1916, Kusatsu 525-0055, Japan \\
$^{3}$The Physics Laboratories, Kyushu institute of 
Technology, Kawazu 680-4, Iizuka 820-8502, Japan
}
\date{\today}
\maketitle
\begin{abstract}
We have revealed that the barrier-tunneling process in non-integrable 
systems is strongly linked to chaos in complex phase space by 
investigating a simple scattering map model.
The semiclassical wavefunction reproduces complicated features of tunneling 
perfectly and it enables us to solve all the reasons why 
those features appear in spite of absence of chaos on the real plane. 
Multi-generation structure of manifolds, which is the manifestation 
of complex-domain homoclinic entanglement created by 
complexified classical dynamics, 
allows a symbolic coding and it is used as a guiding principle to extract 
dominant complex trajectories from all the semiclassical candidates.  

\end{abstract}
\pacs{05.45.Mt, 03.65.Ge, 03.65.Sq, 05.10.-a}

Tunneling phenomenon is peculiar to quantum mechanics and no 
counterparts exist in classical mechanics.
Features of tunneling are nevertheless strongly 
influenced by the underlying classical dynamics
\cite{Wilkinson,Bohigas,Lin,CreaghTunnelReview,ShudoIkeda1,Frischat}.
In particular, chaotic features appearing in tunneling 
have been paid attention to in connection 
with {\it real-domain chaos}\cite{Bohigas,Lin,CreaghTunnelReview}.

A promising approach to see the connection of these two 
opposite concepts is to carry out the complex semiclassical 
analysis, which allows us to describe and interpret the tunneling 
phenomenon in terms of complex classical trajectories \cite{Miller}.
It has been shown that the complex semiclassical theory  
can successfully be applied even in classically chaotic 
systems and the origin of characteristic structures of the wavefunction 
inherent in chaotic systems is explained  
by the complex classical dynamics\cite{ShudoIkeda1}. 
A significant role of almost real-domain homoclinic trajectories 
in the energy barrier tunneling has been pointed out based on the 
trace formula approach\cite{Creagh}. 
Recently, it is found that fringed pattern appears in the wavefunction 
of the two-dimensional barrier tunneling problem
as a result of interference between 
oscillatory Lagrangian manifold\cite{TakahashiYoshimotoIkeda}. 
They have shown a detailed scenario describing how 
such interference emerges in accordance with 
the divergent movement of singularities on the complex $t$-plane. 

In the present paper, we shall report the strong connection 
between the barrier-tunneling process in non-integrable systems 
and the {\it chaos in complex phase space} 
by analyzing a simple scattering map model.  
In particular, it will be shown that even though 
the real-domain classical dynamics exhibits no chaos, i.e. null 
topological entropy, complex-domain chaos can make
tunneling process complicated.  
Moreover our present analysis suggests that  chaotic tunneling can be 
understood in a unified manner from the viewpoint of 
complex-domain chaos not only in case of dynamical but also 
energy-barrier tunneling processes.

\begin{figure}
\begin{center}
\outputfig{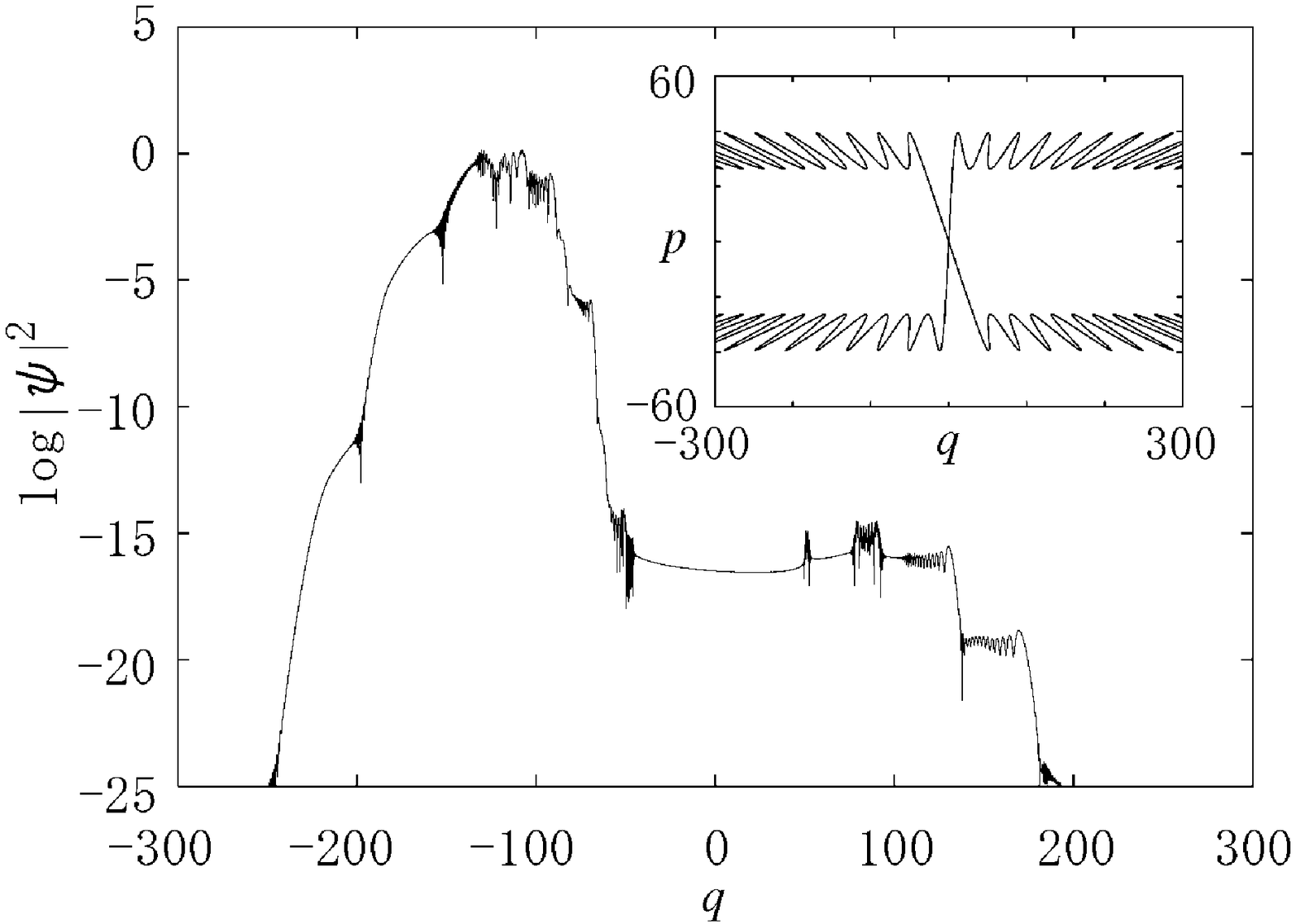}{0.6}{1.0}
\end{center}
\caption{$ \bigl| \langle q | U ^n | \Psi \rangle \bigr| ^2$ for $n=10$ 
calculated quantum-mechanically 
($\hbar \!=\! 1,$  $\sigma \!=\! 10,$ $k \!=\! 500, \gamma \!=\! 
0.005,$  $q_{\alpha} \!=\! -123,$ $p_{\alpha} \!=\! 23$). 
An incident wave packet is set in the side of $q < 0$. 
The center of mass has been already reflected by the 
potential barrier at this time step. (Inset) $ {\cal W}^s(0,0)$ and 
$ {\cal W}^u(0,0)$ merely oscillating without homoclinic entanglement.}
\label{quantum-calculation}
\end{figure}

We first introduce a scattering map model which is described by the 
following Hamiltonian
\begin{mathletters}
\label{hamiltonian}
\begin{eqnarray}
{\cal H}(q,p,t) & = & T(p) \,\, + \,\, V(q) \sum _n \delta (t-n) , \\
T(p) = &p^2 /2& , \quad V(q) = k \exp (-\gamma q^2 ) \quad (k,\gamma > 0). 
\end{eqnarray}
\end{mathletters}
A set of classical equations of motion is given as 
$(q_{j+1},p_{j+1}) = (q_j, + T^{\prime} (p_j), 
p_j - V^{\prime} (q_{j+1}) )$, 
where prime denotes a differentiation with respect to the corresponding 
argument. 
Note that the real-valued dynamics of our scattering map 
does not create chaos in contrast to the maps defined on the bounded phase 
space. This is because the system has only single periodic orbit, 
$(q,p)=(0,0)$, and thus the topological entropy of the system is null.  
As shown in the inset of Fig. \ref{quantum-calculation}, the stable and 
unstable manifolds of the fixed point 
$(0,0)$, ${\cal W}^s(0,0)$ and ${\cal W}^u(0,0)$, 
oscillate without creating homoclinic or heteroclinic intersections.
 Any manifold initially put on the real plane is fully stretched 
 but not folded perfectly so that it leaves away to infinity along  
${\cal W}^u(0,0)$. 

An incident wave packet is put in the asymptotic region. The initial 
kinetic energy is given far less than the potential barrier 
located around the origin. 
The form of initial wave packet is given by
\begin{equation}
\hspace{0pt}
\langle q | \Psi \rangle \, = \, {\cal C}
\exp \left[ - \frac{ (q-q_{\alpha} )^2 }{ 2 \hbar \sigma ^2 } \right] \exp 
\left[ -i \frac{ p_{\alpha} (q_{\alpha} - 2q) }{ 2 \hbar } \right], 
\end{equation}
where ${\cal C}$ is the normalization constant, $\sigma$ is the squeezing 
parameter, and $q_{\alpha},p_{\alpha}$ are configuration and momentum of 
the center of mass, respectively. 

$\bigl| \langle q | U^n | \Psi \rangle \bigr| ^2$ for $n=10$ 
is shown in Fig. \ref{quantum-calculation}, where $U$ denotes the  
unitary operator of 1-step quantum propagation. 
Although the mean energy of the wavepacket is less than the 
barrier height, we can observe various structures   
such as crossover of amplitude, existence of plateau regions,  
erratic oscillation on them, cliffs and so on.  
Similar structures are also found  in case of dynamical tunneling 
in the presence of chaos\cite{ShudoIkeda1},  and we hereafter call such
 characteristic structures ``plateau-cliff structure'' as a 
representative of typical features which are completely absent in  
one-dimensional tunneling.
However, an essential difference of the present situation from the 
previous one is that the plateau-cliff structure 
in the scattering model emerges under the situation where the real-valued 
classical dynamics creates no chaos.

In order to carry out complex semiclassical analysis, 
we first prepare a pair of new canonical variables as
$Q \equiv (q-ip \sigma   ^2 )/(\sqrt{2} \sigma),$ 
$P \equiv (p-iq \sigma ^{-2})/(\sqrt{2} \sigma ^{-1}),$ 
and some notations as $Q_0 \equiv Q(q_0,p_0)$, $P_0 \equiv P(q_0,p_0)$, 
$Q_{\alpha} \equiv Q(q_{\alpha},p_{\alpha})$, 
$P_{\alpha} \equiv P(q_{\alpha},p_{\alpha})$.
The {\it n}-step quantum propagator is  
represented as the {\it n}-fold  multiple integral, and 
the saddle point evaluation is applied to 
yield the classical equations. 
Semiclassical Van Vleck's formula for {\it n}-step 
wavefunction takes a form as 

\begin{equation}\label{semiclassical_sum}
\langle q_n | U^n | \Psi \rangle \,\, \approx \,\, {\cal A} 
\sum_{cl.orb.} 
\Bigl| 
\frac{\partial ^2 W}{\partial q_n \partial P_0 } \Bigr| 
^{\frac{1}{2}} \exp \frac{i}{\hbar} \Big[ {\cal S} - \frac{\phi}{2} \Big],
\end{equation}

\noindent
where the sum is over complex classical orbits satisfying the boundary 
condition, Im\, $q_n = 0$, which is necessary since we here observe our 
wavefunction as a function of $q_n$. 
$ {\cal A} $ denotes the normalization factor, and ${\cal 
S} , \phi$ are the action and the Maslov index of each contributing orbit 
respectively. {\it W} is the generating function which gives canonical 
transformations such 
that $ \partial W / \partial q_n = p_n , \,\, \partial W / 
\partial P_0 = -Q_0$ . 

In order to represent the complex orbits which can contribute to 
the semiclassical propagator (\ref{semiclassical_sum}), 
we introduce a variable, $\Delta Q_0 \equiv Q_0 \, - \, Q_{\alpha}$.  
Using $\Delta Q_0$ the contributing orbits 
in (\ref{semiclassical_sum}) are represented as a set of initial conditions 
${\cal M}_n \equiv \left\{ \Delta Q_0 \,\, | \,\,
\Delta Q_0 \in \hbox{\bf C}, \,
\hbox{Im} \, q_n (Q_{\alpha} \! \! +  \! \Delta Q_0 , \, P_{\alpha}) 
\, = \, 0 \right\}$.

\begin{figure}
\begin{center}
\outputfig{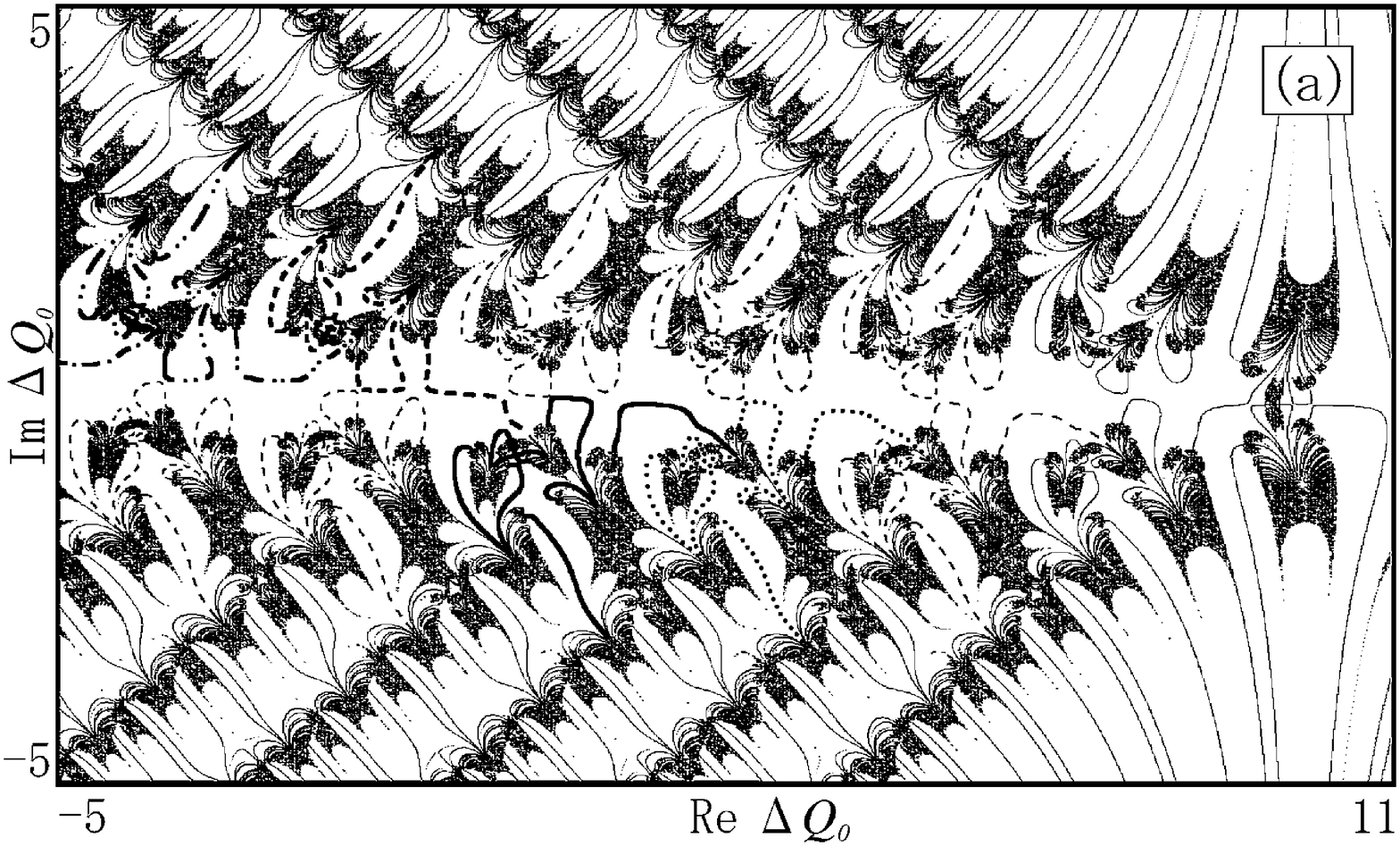}{0.6}{0.9}
\outputfig{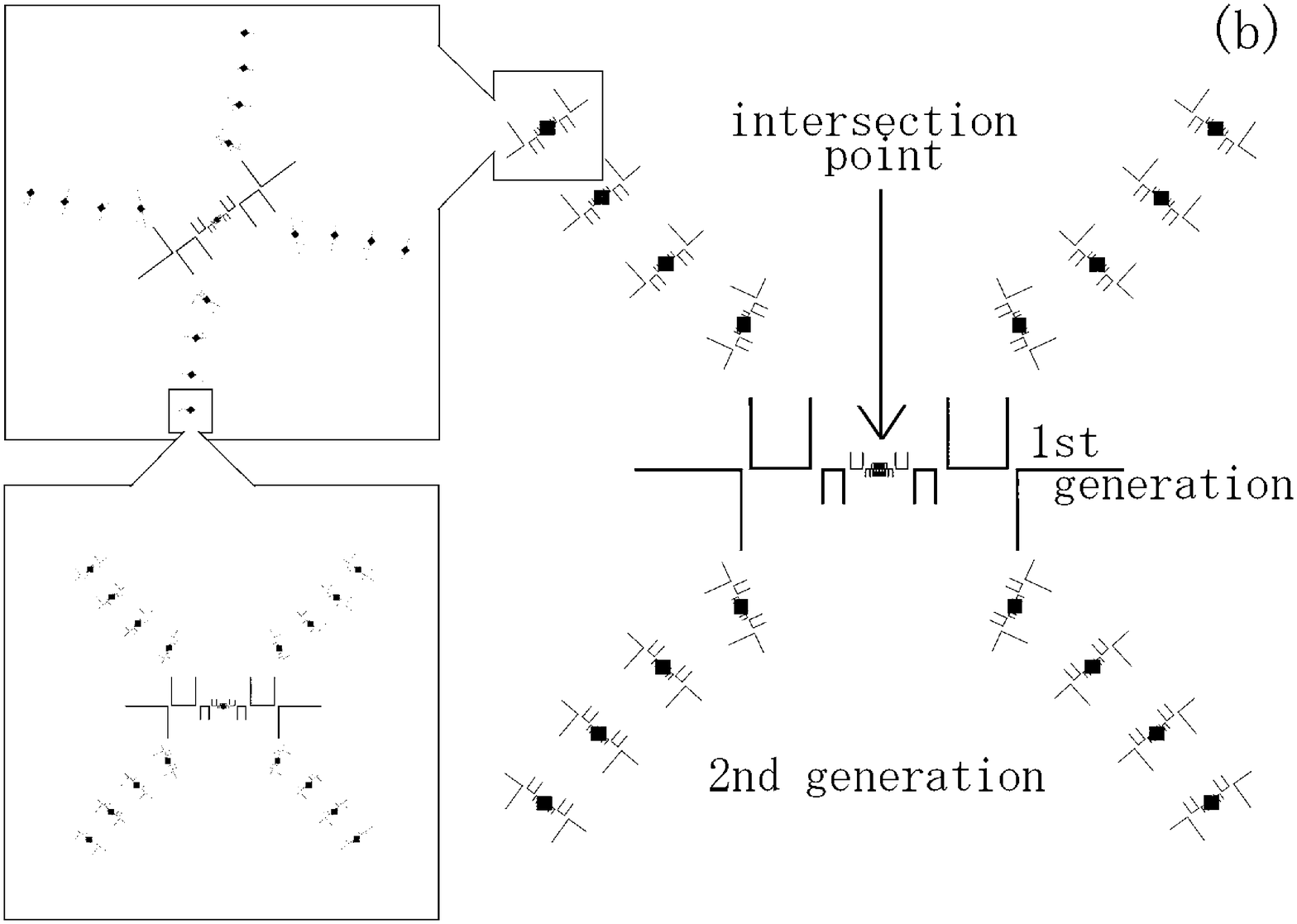}{0.54}{0.81}
\end{center}
\caption{
(a) ${\cal M}_n$-set for $n = 10$. 
Those branches which we finally take 
into account for the reproduction of the wavefunction
are represented by the line styles such as 
dash, bold dash, bold solid, bold dot, bold dash dot 
and bold dash dot dot. Each line style corresponds to 
the one in Fig. \ref{semiclassical-wavefunction}.
(b) Schematic representation for MGS. Blown-up figures of the 
second and third generations display the self similarity of MGS. 
Solid squares represent the intersection points with ${\cal W}^s(0,0)$.
}
\label{M-set}
\end{figure}

A typical pattern of ${\cal M}_n$-set is shown in Fig. \ref{M-set}(a). 
Each string, which we call the branch hereafter, 
represents an individual classical orbit appearing the 
semiclassical sum (5) and a single string covers 
the whole range $(-\infty , +\infty)$
of the final $q_n$. 

Though the morphology of the set is quite elusive, when sufficiently 
blowing up any local area with lots of branches accumulated, one can find 
the self-similar structure 
schematically demonstrated in Fig. \ref{M-set}(b). At the center, 
a chain-shaped structure is developed in the horizontal direction. 
A sequence of chains with finer scale are arranged in both sides of 
the previous chain, and does the same around any of those smaller chains, 
and so on. 
Then it may be natural to assign a notion of the {\it generation} to 
each chain as shown in the figure. Higher generations 
sprout from lower generations almost vertically, and 
we call these self-similar configurations the multi-generation 
structure(MGS). 

We can discuss significant branches embedded in the hierarchical structure, 
which determine the semiclassical wavefunction (\ref{semiclassical_sum}). 
The amplitude of an individual contribution 
in the semiclassical sum is almost governed by Im\,${\cal S}$, 
an imaginary action, which implies that 
the orbit whose imaginary part 
is the smallest survives as the final semiclassical contribution. 
Furthermore, the amount of Im\,${\cal S}$ is roughly estimated 
by how deep each trajectory passes through the complex domain. 
This means that the history of 
each trajectory in complex domain is a primary factor concerning its weight 
in the semiclassical sum. 
It implies that the semiclassical wavefunction could not be reproduced 
untill one finds how those orbits with the smallest imaginary actions 
behave, since it is almost impossible to extract significant 
trajectories out of a huge number of branches without any 
guiding principle. 

To this end, we begin with the precise 
definition of MGS, which is given by considering 
the {\it stable and unstable manifolds in complex phase space}, in
particular analytical continuation of the stable manifold ${\cal W}^s(0,0)$. 
In order to connect MGS with ${\cal W}^s(0,0)$ in complex space, 
we first introduce the normalized coordinate on ${\cal W}^s(0,0)$. 
Let $\Phi$ be the conjugation map from $\hbox{\bf C}$ to ${\cal W}^s(0,0)$, 
satisfying the relation $\tilde{\cal F}(\xi) \equiv \Phi^{-1}\!\cdot\! 
{\cal F}\!\cdot\!\Phi (\xi) = {\lambda}^{-1} \xi$, where $\lambda$ denotes 
the largest eigenvalue with respect to the saddle point $(0,0)$
\cite{Gelfreich1}.  

\begin{figure}
\begin{center}
\outputfig{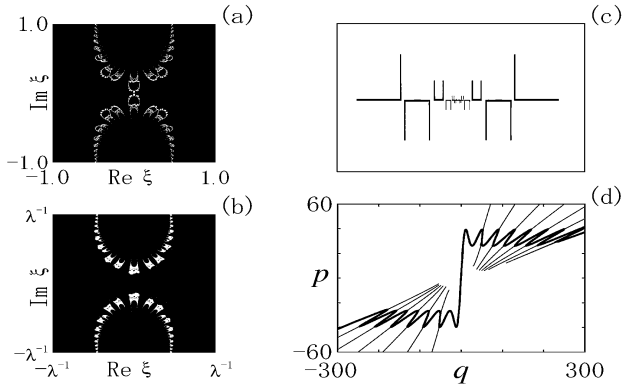}{0.6}{1.0}
\end{center}
\caption{
(a) The intersection points belonging to the 1st, 2nd or 3rd 
generation plotted on ${\cal W}^s(0,0)$.
(b) Magnification of (a), displaying the 1st and 2nd 
generations.
(c) Chain-shaped manifolds in ${\cal M}_n$-set drawn 
schematically. 
(d) The final strings of those displayed in (c) 
after 10 iterations of our map projected in 
real phase space. The hatched part in (c) develops 
to the bold curves, which almost coincide with 
the real-domain unstable manifold, 
${\cal W}^u(0,0)\cap\hbox{\bf R}^2$.
}
\label{on_stable_mfd}
\label{MandLset}
\end{figure}

The normalized coordinate $\xi$ is used in Fig. \ref{on_stable_mfd}(a) 
and (b) to represent the intersection between 
${\cal W}^s(0,0)$ and ${\cal I}$, the initial value plane 
consisting of the whole $(q,p)$'s corresponding to 
$\{ \Delta Q_0 \, | \, \Delta Q_0 \in \hbox{\bf C} \}$. 
Comparing both figures, one can recognize 
the intersection pattern has self-similarity. 
The intersection between ${\cal W}^u(0,0)$ and ${\cal W}^s(0,0)$ 
shows almost the same self-similar pattern, 
which implies the {\it homoclinic entanglement} 
of the stable manifold in complex phase space
\cite{OnishiShudoIkedaTakahashi1}. 
This means that 
null topological entropy in the real domain 
does not exclude existence of chaos in the complex domain. 

The generation can be assigned to 
each point on ${\cal I}\,\cap\, {\cal W}^s(0,0)$. 
Suppose $r$ be the minimum distance on $\xi$-plane from $(0,0)$ to 
intersection points $\Phi^{-1}({\cal I}\,\cap\, {\cal W}^s(0,0))$, 
and let $D \equiv \{ \xi \in \hbox{\bf C}\, | \, |\xi| \le r \}$. 
Here the normalized coordinate plane is decomposed into 
disjointed annulus as, 
\begin{eqnarray}\label{definition_of_generation}
\tilde{\cal F}^m(D\backslash \tilde{\cal F}(D)) \cap 
\tilde{\cal F}^n(D\backslash \tilde{\cal F}(D)) \quad & = & \quad\phi 
\quad (m \neq n) \\
\bigsqcup_{n\in\hbox{\bf Z}}
\tilde{\cal F}^n(D\backslash \tilde{\cal F}(D)) \quad & = & \quad\hbox{\bf C}
\end{eqnarray}
Then each annulus $\tilde{\cal F}^n (D\backslash \tilde{\cal F}(D))$ 
plays a role specifying the individual generation. 
More precisely, if a point of ${\cal I}\,\cap\, {\cal W}^s(0,0)$ is 
contained in 
$\Phi \cdot \tilde{\cal F}^{-n}(D\backslash \tilde{\cal F}(D))$, 
we say the point belongs to the $n$-th generation. 

Next we show each point of ${\cal I}\,\cap\, {\cal W}^s(0,0)$ 
can be associated with a single chain structure on ${\cal M}_n$-set 
which is shown as the hatched zone in Fig. \ref{MandLset}(c).  
The final strings of those displayed in the figure 
after 10 iterations of our map are projected 
in real phase space in Fig. \ref{MandLset}(d). 
The hatched part in Fig. \ref{MandLset}(c) develops to the bold 
curves in Fig. \ref{MandLset}(d),  
which almost coincide with the real-domain unstable manifold, 
${\cal W}^u(0,0)\cap\hbox{\bf R}^2$. 

This behavior is not limited to a particular chain-shaped structure, 
but every chain structure in ${\cal M}_n$-set becomes close to  
${\cal W}^u(0,0)\cap\hbox{\bf R}^2$. 
Chain-shaped structure is always created around each point of
${\cal I}\,\cap\,{\cal W}^s(0,0)$,
the reason of which can be explained by describing 
the process of time evolution
of a tiny domain containing a point of ${\cal I}\,\cap\, {\cal W}^s(0,0)$. 
Such a domain is guided to the real phase space 
by ${\cal W}^s(0,0)$ and then smeared over ${\cal W}^u(0,0)$.
Details will be reported in \cite{OnishiShudoIkedaTakahashi1}.
In this way, we can find one-to-one relation between a single 
chain structure in ${\cal M}_n$-set and an intersection 
point of ${\cal I}\,\cap\,{\cal W}^s (0,0)$.  
Since, as explained above, 
the latter forms the self-similar or generation structure 
reflecting the homoclinic entanglement of the stable manifold, 
MGS of ${\cal M}_n$-set is controlled by that of the stable manifold in 
complex phase space.

Since we have defined the generation in such a way, 
the orbits belonging to higher generations usually gain larger Im S, 
since, taking roundabout ways in the complex chaotic region, 
they approach the origin later than those of lower generations. 
In order to extract significant trajectories out of them, 
it is essential to use symbolic dynamics in complex phase space.

In the present case, a nice symbolic space can be organized as 
the union of a single-element set $\left\{ O \right\}$ and 
a direct product of the sign of Re\,$q$, Im\,$q$ and a multiplicity 
index $\nu \in {\bf N}$. The symbol `$O$' describes monotonic 
convergence to the origin $(0,0)$, and the index $\nu$ appears 
as a reflection of the transcendental property 
of the potential function such that our initial-value problem has 
infinite number of solutions. 
Since frequent flipping of the sign or large modulus of $q$-component 
which is represented as large $\nu$, 
result in large imaginary actions, our final principle turns out to 
extract the sequences of symbols out of MGS which represent those orbits with 
no sign flipping and minimal $q$-components with $\nu = 1$. 
It considerably reduces our task in searching significant complex 
orbits only to linear dependence of the time step, otherwise 
exponentially diverges. 
More detailed explanations to construct symbolic dynamics 
in the complex-domain chaos will be reported 
elsewhere \cite{OnishiShudoIkedaTakahashi1}. 
In particular, complex H\'enon map is analyzed to elucidate 
the relation between the MGS and Julia set\cite{ShudoIshiiIkeda1}. 

\begin{figure}
\begin{center}
\outputfig{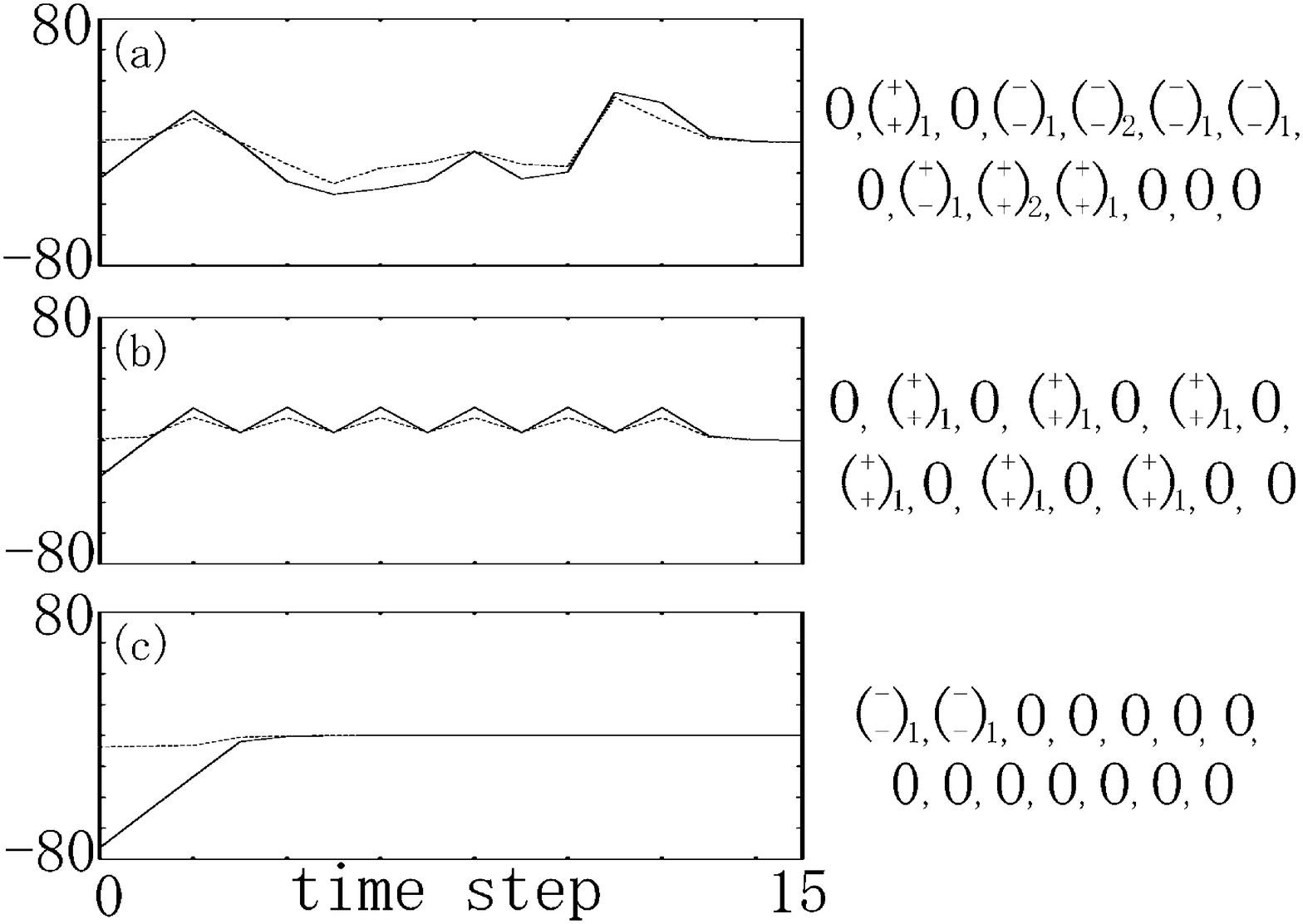}{0.6}{1.0}
\end{center}
\caption{
Behaviors of intersection points in MGS. Solid lines as Re 
$q$, and broken lines as Im $q$. The sequences of symbols 
coding them also attached. These are the trajectories 
(a) showing stochastic motions,
(b) temporarily attracted by a complex period-2 orbit, 
(c) approaching real phase space monotonically. 
}
\label{Mset_and_stable_foliation}
\label{higher-generations}
\end{figure}

Fig. \ref{higher-generations} displays the behaviors 
of intersection points in MGS as a function of time, together  
with the coding sequences.
Fig. \ref{higher-generations}(a) shows a typical behavior 
such that both Re\,$q$ and Im\,$q$ oscillate in an erratic manner 
for some initial time steps and eventually approach the origin.
Fig. \ref{higher-generations}(b) shows such an orbit trapped 
by complex period-2 orbits as evidence of regular behavior embedded 
in MGS.
Such itinerating or oscillating complex orbits contribute
in principle, however, their amplitudes are 
much smaller than those of orbits with 
Re\,$q$ and Im\,$q$ decreasing monotonically to zero 
as shown in Fig. \ref{Mset_and_stable_foliation}(c). 

\begin{figure}
\begin{center}
\outputfig{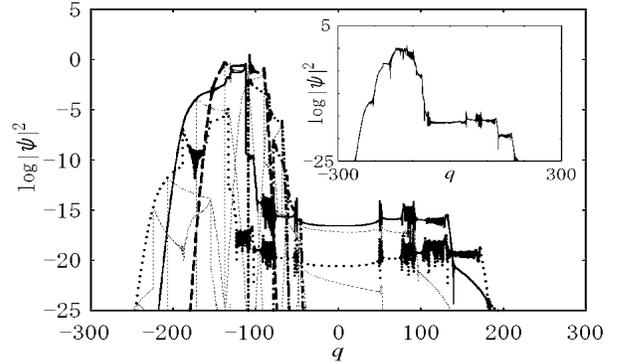}{0.6}{1.0}
\end{center}
\caption{
The semiclassical wavefunction for 10 time steps 
resolved into contributions from different generations. 
Each component corresponds to the branches with 
the same line styles in Fig. \ref{M-set}(a). 
The thin dashed lines represent those branches 
which are dominated, in each 
coordinate $q$, by some other branches. 
(Inset) 
The semiclassical wavefunction finally obtained 
by the superposition of these contributions.
}  
\label{semiclassical-wavefunction}
\end{figure}

The semiclassical wavefunction evaluated finally takes 
the form in Fig. \ref{semiclassical-wavefunction}. 
The agreement with the quantal one shown in Fig. \ref{quantum-calculation} 
is excellent. The origin of plateau-cliff structure is understood by 
resolving the superposed semiclassical function into contributions 
from individual generations. The 
plateau-cliff structure is created because each semiclassical component 
itself has a flat plateau accompanied by sharp drop and interference 
between branches forming the generation give rise to erratic oscillation 
on the plateau. Discontinuity of amplitude, as found in $q\approx -168,-207,
-224,$ etc. are caused by the Stokes phenomenon which is inevitable in 
the saddle point method, and is treated in an appropriate way
\cite{ShudoIkeda2}. 

It should be stressed that the most crucial step in our semiclassical 
analysis is to decode embedded information in MGS which reflects 
complex homoclinic entanglement. 
This means that emergence of the entanglement in complex phase 
space is an essential ingredient 
in our description. 
We think it is a quite general event in chaotic systems, 
and thus so-called plateau-cliff structure, which is typically observed 
pattern of tunneling in the presence of chaos\cite{ShudoIkeda1}, 
must appear as a manifestation of such complex structures. 
We can therefore expect that, whether energetic or dynamical, 
there is a common semiclassical mechanism of the tunneling phenomena  
in chaotic map systems, which should be attributed to the 
structure of entanglement, typically observed as MGS.

\end{document}